\documentclass{article}
\usepackage{epsf,epsfig,pstricks}
\title{The importance of the observer in science}
\author{Russell K. Standish\\School Mathematics, UNSW\\
R.Standish@unsw.edu.au\\http://parallel.hpc.unsw.edu.au}

\begin{document}

\maketitle

\begin{abstract}
  The concept of {\em complexity} (as a quantity) has been plagued by
  numerous contradictory and confusing definitions. By explicitly
  recognising a role for the observer of a system, an observer that
  attaches meaning to data about the system, these contradictions can
  be resolved, and the numerous complexity measures that have been
  proposed can be seen as cases where different observers are
  relevant, and/or being proxy measures that loosely scale with
  complexity, but are easy to compute from the available data. Much of
  the epistemic confusion in the subject can be squarely placed at
  science's tradition of removing the observer from the description in
  order to guarantee {\em objectivity}.
  
  Explicitly acknowledging the role of the observer helps untangle
  other confused subject areas. {\em Emergence} is a topic about which
  much ink has been spilt, but it can be understand easily as an
  irreducibility between description space and meaning space. Quantum
  Mechanics can also be understood as a theory of observation. The
  success in explaining quantum mechanics, leads one to conjecture
  that all of physics may be reducible to properties of the observer.

  And indeed, what are the necessary (as opposed to contingent)
  properties of an observer? This requires a full theory of
  consciousness, from which we are a long way from obtaining. However
  where progress does appear to have been made, e.g. Daniel Dennett's
  {\em Consciousness Explained}, a recurring theme of self-observation
  is a crucial ingredient.
\end{abstract}

\section{Introduction}
I have set myself the ``humble'' task of understanding how evolution
leads to continuous generation of complexity and novelty. To
circumscribe certain unproductive lines of argument, I take as given
that biological evolution proceeds through a perfectly mechanistic
process, that there is no supernatural intervention. What remains is the
task of reverse engineering the evolutionary process.

Since the 1970s, various {\em evolutionary algorithms}\cite{Fogel95}
have been proposed and implemented on computers. None of these
algorithms have clearly demonstrated open-ended creativity, or growth
of complexity\cite{Bedau-etal00}. In fact, the very notion of
complexity is muddy\cite{Edmonds99}, which is caused by the
traditional scientific notion of objectivity as removing the observer
from the description\cite{Standish01a}.

Complexity turns out to be intrinsically observer dependent. This is
not the disaster many scientists might fear, however, as in all
applicable cases there will be a natural choice of observer. Including
the observer into the picture actually simplifies and unifies the
disparate notions of complexity that have been proposed.

Complex systems theory is not the only area where including the
observer makes theories comprehensible. There are three independent
derivations of the fundamental postulates of quantum mechanics, based
on assumed properties of the observer. Each of these has a slightly
different starting point: Bruno Marchal assumes that an observer is
equivalent to some unspecified computer program\cite{Marchal98}, Roy Frieden starts
from Fisher information theory\cite{Frieden98}, a branch of statistics that predicts
the observational error for an ideal observer, and my own derivation
starts from an all universes ensemble with zero information, a
psychological phenomenon of time (implicitly assumed in both Marchal
and Frieden's approaches) and projection as a model of measurement
(observed outcomes are selected from the set of possible observation
according to a probability distribution)\cite{Standish00a}.

\section{Complexity as a quantity}

We have an intuitive notion of complexity as a quantity; we often
speak of something being more or less complex than something
else. However, capturing what we mean by complexity in a formal way
has proved far more difficult, than other more familiar quantities we
use, such as length, area and mass. 

In these more conventional cases, the quantities in question prove to
be decomposable in a linear way, ie a 5cm length can be broken into 5
equal parts 1 cm long; and they can also be directly compared --- a
mass can be compared with a standard mass by comparing the weights of
the two objects on a balance.

However, complexity is not like that. Cutting an object in half does
not leave you with two objects having half the complexity overall. Nor
can you easily compare the complexity of two objects, say an apple and
an orange, in the same way you can compare their masses. However, the
earliest attempts at deriving a measure took this approach. 

The simplest such measure is the {\em number of parts} definition. A
car is more complex than a bicycle, because it contains more
parts. However, a pile of sand contains an enormous number of parts
(each grain of sand), yet it is not so complex since each grain of
sand is conceptually the same, and the order of the grains in the pile
is not important. Another definition used is the {\em number of
distinct parts}, which partially circumvents this problem. The problem
with this idea is that a shopping list and a Shakespearian play will
end up having the same complexity, since it is constructed from the
same set of parts (the 26 letters of the alphabet --- assuming the
shopping list includes items like zucchini, wax and quince, of
course). An even bigger problem is to define precisely what one means
by ``part''. This is an example of the {\em context dependence} of
complexity, which we'll explore further later.

Bonner\cite{Bonner88} and McShea\cite{McShea96} have used these
(organism size, number of cell types) and other {\em proxy} complexity
measures to analyse complexity trends in evolution. They argue that
all these measures trend in the same way when figures are available
for the same organism, hence are indicative of an underlying organism
complexity value. This approach is of most value when analysing trends
within a single phylogenetic line, such as the diversification of
trilobytes.

\subsection{Information as Complexity}\label{info-complexity}

The single simplest unifying concept that covers all of the preceding
considerations is {\em information}. The more information required to
specify a system, the more complex it is. A sandpile is simple,
because the only information required is that it is made of sand
grains (each considered to be identical, even if they aren't in
reality), and the total number of grains in the pile. However, a
typical motorcar requires a whole book of blueprints in its
specification.

Information theory began in the work of Shannon in the 1940s, who was
concerned with the practical problem of ensuring reliable transmission
of messages. Every possible message has a certain probability of
occurring. The less likely a message is, the more information it
imparts to the listener of that message. The precise relationship is
given by a logarithm:
\begin{equation}\label{shannon}
I = -\log_2 p
\end{equation}
where $p$ is the probability of the message, and $I$ is the
information it contains for the listener. The base of the logarithm
determines what units information is measured in --- base 2 means the
information is expressed in {\em bits}. Base 256 could be used to
express the result in {\em bytes}, and is of course equivalent to
dividing equation (\ref{shannon}) by 8.

\begin{figure}
\psset{unit=0.8cm}
\begin{pspicture}(0,1)(15,9)
\rput[l](0,5){\epsfxsize=8\psxunit\epsfbox{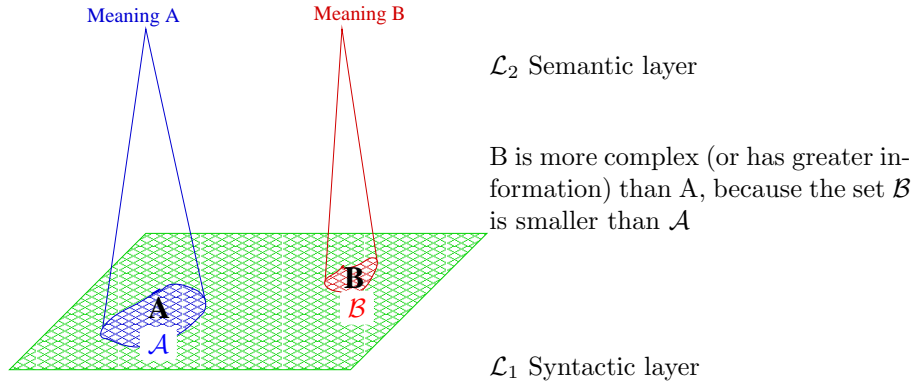}}
\rput(2.5,2.4){{\blue\psframebox*[framearc=.3]{${\cal A}$}}}
\rput(5.8,3){{\red\psframebox*[framearc=.3]{${\cal B}$}}}
\rput[l](8,2){${\cal L}_1$ Syntactic layer}
\rput[l](8,7){${\cal L}_2$ Semantic layer}
\rput[l](8,5){\parbox{7\psxunit}{B is more complex (or has greater
    information) than A, because the set ${\cal B}$ is smaller
  than ${\cal A}$}}
\end{pspicture}
\caption{Diagram showing the syntactic and semantic spaces. Two
  different messages, having meanings A and B, can each be coded in
  many equivalent ways in syntactic space, represented by the sets
  ${\cal A}$ and ${\cal B}$. The information or complexity of the
  messages is related to the size it occupies in syntactic space by
  formula (\ref{shannon})}
\label{syn-sem}
\end{figure}

Writing the probability of a symbol $x$ drawn from an alphabet $A$
appears in message $s$ as $p(x)$, equation (\ref{shannon}) can be
approximated:
\begin{equation} 
I(s) \approx \sum_{i=1}^n\sum_{x_i\in A} p(x_i)\log_2 p(x_i).
\end{equation}
Shannon, of course, was not so interested in the semantic content of
the message (ie its meaning), rather in the task of information
transmission so the probability distributions $p(x)$ were simply
those of the symbols for the language in question, eg English, for
which the letter `e' is considerably more likely than the letter `q'.
This equation can be refined by considering possible pairs of letters,
then possible triplets, in the limit converging on the minimum amount
of information required to be transmitted in order for the message to
be reconstructed in its original form. That this value may be
considerably less that just sending the original message in its
entirety is the basis of compression algorithms, such as those
employed by the well-known {\em gzip} or {\em PKzip} (aka WinZip)
programs.

The issue of semantic content discouraged a lot of people of applying
this formalism to complexity measures. The problem is that a message
written in English will mean something to a native English speaker,
but be total gibberish to someone brought up in the Amazon jungle with
no contact with the English speaking world. The information content of
the message depends on exactly who the listener is! Whilst this
context dependence appears to make the whole enterprise hopeless, it
is in fact a feature of all of the measures discussed so far. When
counting the number of parts in a system, one must make a decision as
to what exactly constitutes a part, which is invariably somewhat
subjective, and needs to be decided by consensus or convention by the
parties involved in the discussion. Think of the problems in trying
decide whether a group of animals is one species of two, or which
genus they belong to. The same issue arises with the characterisation
of the system by a network. When is a relationship considered a graph
edge, when often every component is connected to every other
part in varying degrees.

However, in many situations, there appears to be an obvious way of
partitioning the system, or categorising it. In such a case, where two
observers agree on the same way of interpreting a system, then they
can agree on the complexity that system has. If there is no agreement
on how to perform this categorisation, then complexity is meaningless

To formalise complexity then, assume as given a classifier system
that can categorise descriptions into equivalence classes. Clearly,
humans are very good at this --- they're able to recognise patterns
even in almost completely random data. Rorschach plots are random ink
plots that are interpreted by viewers as a variety of meaningful
images. However, a human classifier system is not the only
possibility. Another is the classification of programs executed by a
computer by what output they produce. Technically, in these
discussions, researchers use a {\em Universal Turing Machine} (UTM),
an abstract model of a computer.

Consider then the set of possible binary strings, which can fed into a
UTM $U$ as a program. Some of these programs cause $U$ to produce some
output then halt. Others will continue executing forever. In
principle, it is impossible to determine generally if a program will
halt or continue on indefinitely. This is the so called {\em halting
problem}. Now consider a program $p$ that causes the UTM to
output a specific string $s$ and then halt. Since the UTM halts after
a certain number of instructions executed (denoted $\ell(p)$) 
the same result is produced by feeding in any string starting with the same
$\ell(p)$ bits.  If the strings have equal chance of being chosen
({\em uniform measure}), then the proportion of strings starting with the
same initial $\ell(p)$ bits is $2^{-\ell(p)}$. This leads to the {\em
universal prior} distribution over descriptions $s$, also known as the
Solomonoff-Levin distribution:
\begin{equation}\label{universal prior}
P(s) = \int_{U(p)=s}2^{-\ell(p)} dp
\end{equation}

The complexity (or information content) of the description is given by
equation (\ref{shannon}), or simply the logarithm of (\ref{universal
prior}). In the case of an arbitrary classifier system, the
complexity is given by the negative logarithm of the equivalence class size
\begin{equation}\label{complexity}
C(x) = \lim_{s\rightarrow\infty} s\log_2 N - \log_2 \omega(s,x)
\end{equation}
where $N$ is the size of the alphabet used to encode the description
and $\omega(s,x)$ is the number of equivalent descriptions to $x$ of
size $s$ or less.

It turns out that the probability $P(s)$ in equation (\ref{universal
prior}) is dominated by the shortest program, namely 
\begin{displaymath}
K(s) \leq \log_2 P(s) \leq K(s) + C
\end{displaymath}
where $C$ is a constant independent of the description $s$. $K(s)$ is
the length of the shortest program $p$ that causes $U$ to output $s$
and halt, and is called the {\em Kolmogorov Complexity} or {\em
Algorithmic Complexity}.

An interesting difference between Kolmogorov Complexity, and the
general complexity based on human observers can be seen by considering
the case of random strings. {\em Random}, as used in algorithmic
information theory, means that no shorter algorithm can be found to
produce a string than simply saying ``print \ldots'', where the
\ldots{} is a literal representation of the string. The Kolmogorov
complexity of a random string is high, at least as high as the length
of the string itself. However, a human observer simply sees a random
string as a jumble of letters, much the same as any other random
string. In this latter case, the equivalence class of random strings
is very large, close to probability one, so the perceived complexity
is small. Thus the human classifier defines an example of what
Gell-Mann calls {\em effective complexity}\cite{Gell-Mann94}, which
measures the length of a concise description of the regular parts of
the description, ignoring the random components.

\section{Complexity as a quality --- Emergence}

It is often thought that {\em complex systems} are a separate category
of systems to {\em simple systems}. So what is it that distinguishes a
complex system, such as a living organism, or an economy, from a
simple system, such as a pair of pliers? This question is related to
the notorious question of {\em What is Life?}, however may have a
simpler answer, since not all complex systems are living, or even
associated with living systems.

Consider the concept of {\em emergence}. We intuitively recognise
emergence as patterns arising out of the interactions of the
components in a system, but not implicit in the components
themselves. Examples include the formation of hurricanes from pressure
gradients in the atmosphere, crashes in stock markets, flocking
behaviour of many types of animals and of course, life itself.

Let us consider a couple of simple illustrative examples, that are
well known and understood. The first is the {\em ideal gas}, a model
gas made up of large numbers of non-interacting point particles
obeying Newton's laws of motion. A {\em thermodynamic} description of
the gas is obtained by averaging: 
\begin{description}
\item[temperature ($T$)] is the average kinetic energy of the particles;
\item[pressure ($P$)] is the average force applied to a unit area of the
boundary by the particles colliding with it;
\item[density ($\rho$)] is the average mass of particles in a unit volume;
\end{description}
The {\em ideal gas law} is simply a reflection of the underlying laws
of motion, averaged over all the particles:
\begin{displaymath}
P\rho\propto T
\end{displaymath}
The thermodynamic state is characterised by the two parameters $T$ and
$\rho$. The so-called {\em first law of thermodynamics} is simply a
statement of conservation of energy and matter, in average form.

An entirely different quantity enters the picture in the form of {\em
entropy}. Consider discretising the underlying phase-space into cubes
of size $h^N$, ($N$ being the number of particles) and then counting
the number of such cubes having temperature $T$ and density $\rho$,
$\omega(T,\rho,N)$. The entropy of the system is given by
\begin{equation}
S(T,\rho,N)=k_B \ln \omega(T,\rho,N)
\end{equation}
where $k_B$ is a conversion constant that expresses entropy in units
of Joules per Kelvin.  One can immediately see the connection between
complexity (eq. \ref{complexity}) and entropy. Readers familiar with
quantum mechanics will recognise $h$ as being an analogue of Planck's
constant. However, the ideal gas is not a quantum system, and as
$h\rightarrow0$, entropy diverges! It turns out that in the
thermodynamic limit ($N\rightarrow\infty$), the per-particle entropy $S/N$
is independent of the size of $h$.

The {\em second law of thermodynamics} is a recognition of the fact
that the system is more likely to move to a state occupying a larger
region of phase space, than a smaller region of phase space, namely
that $\omega(T,\rho,N)$ must increase in time. Correspondingly entropy
must also increase (or remain constant) over time. This is a
probabilistic statement that only becomes exact in the thermodynamic
limit. At the syntactic, or specification level of description (ie
Newton's laws of motion), the system is perfectly {\em reversible} (we
can recover the system's initial state by merely reversing the
velocities of all the particles), yet at the semantic (thermodynamic)
level, the system is {\em irreversible} (entropy can only increase, never
decrease).

The property of irreversibility is an {\em emergent} property of the
ideal gas, as it is not {\em entailed} by the underlying
specification. It comes about because of the additional identification
of the thermodynamic state, namely the set of all micro-states
possessing the same temperature and density. This is additional
information to what is contained in the microscopic description, and that
entails the second law.

The second example I'd like to raise (but not analyse in such great
depth) is the well known {\em Game of Life}\cite{Conway82}, introduced by John
Conway, and popularised in Martin Gardiner's {\em Mathematical
Recreations} column in 1970. This is a 2 dimensional {\em cellular
automaton} (2D grid of cells), where each cell can be one of two
states. Dynamics on the system is imposed by the rule that the state
of a cell depends on the values of its immediate neighbours at the
previous time step.

Upon running the Game of Life, one immediately recognises a huge
menagery of emergent objects, such as blocks, blinkers and
gliders. Take gliders for example. This is a pattern that moves
diagonally through the grid. The human observer recognises this
pattern, and can use it to {\em predict} the behaviour of the system
with less effort than simulating the full cellular automaton. It is a
{\em model} of the system. However, the concept of a glider is not
{\em entailed} by the CA specification, which contains only states and
transition rules. It requires the additional identification of a
pattern by the observer.

This leads to a general formulation of {\em emergence}. Consider a
system specified in a language ${\cal L}_1$, which can be called the
specification, or {\em syntactic} layer (see figure \ref{syn-sem}). If
one accepts the principle of {\em reduction}, all systems can
ultimately be specified the common language of the theoretical physics
of elementary particles. However, an often believed corollary of
reduction is that this specification encodes all there is to know
about the system. The above two examples shows this corollary to be
manifestly false. Many systems exhibit one or more {\em good models},
in another language ${\cal L}_2$, which can be called the {\em semantic
layer}. The system's specification does not entail completely the
behaviour of the semantic model, since the latter also depends on
specific identifications made by the observer. In such a case, we say
that properties of the semantic model is emergent with respect to the
syntactic specification.

The concept of ``good'' model deserves further discussion. In our
previous two examples, neither the thermodynamic model, nor the glider
model can be said to perfectly capture the behaviour of the
system. For example, the second law of thermodynamics only holds in
the thermodynamic limit --- entropy may occasionally decrease in finite
sized systems. A model based on gliders cannot predict what happens
when two gliders collide. However, in both of these cases, the
semantic model is cheap to evaluate, relative to simulating the full
system specification. This makes the model ``good'' or ``useful'' to
the observer. We don't prescribe here exactly how to generate good
models here, but simply note that in all cases of recognised emergence, the
observer has defined a least one semantic and a syntactic model of the
system, and that these models are fundamentally
incommensurate. Systems exhibiting emergence in this precise sense can
be called {\em complex}.

\section{An Ontology of Bitstrings}

Consider a program of explaining the world of appearances, of
explaining Kantian {\em phenomena}. Kant distinguishes two categories of
being --- the {\em phenomenon}, or thing as it appears, and the {\em
  noumenon}, or thing as it is. If phenomena can be explained in a
closed manner, ie without appeal to an underlying noumenon, then
perhaps the noumenon can be dispatched in the manner of Laplace's
reply to Napoleon Bonaparte: {\em Je n'ai besoin de cet hypoth\`ese}.

Phenomena (appearance) arises through the registering of data by the
conscious observer, through the attachment of meaning to raw data.
This is the situation captured in Fig. \ref{syn-sem}. Suppose,
therefore that all possible bitstrings exist, in uniform measure. By
virtue of equation (\ref{complexity}), the information content of this
complete set is precisely zero, independent of the observer. Such an
assumption is a minimal ontology, on a par with assuming that
nothing exists. One could also paraphrase it as saying {\em no
  constraints exist}. In deference to the influence of Plato, such an
all encompassing object is often termed a {\em Plenitude} or
sometimes {\em Platonia}. I use the indefinite article here, as
alternatives do exist: eg {\em all logically consistent systems}, {\em
  all mathematical systems}\cite{Tegmark98}, {\em all possible
  computations}\cite{Schmidhuber00}. 

Equation (\ref{shannon}) implies that the phenomena observed should be
more likely to be simple than complex, leading to an explanation of
the value of Occam's Razor: that things should not be multiplied
unnecessarily\cite{Standish00a}. 

Furthermore, by supposing that the observer is parameterised by an
ordered set (called (psychological) {\em time}) such that only a
finite number (but increasing as a function of time) of bits of the
bitstring are meaningful, and also that observers' interpretations are
robust in the presence of noise (evolutionarily speaking, it is not
a good idea to be fooled by lions in camouflage), we get a solution to
the problem of induction. The world will, by and large, continue to
follow the rules it has followed previously\cite{Standish00a}.

Some further elaboration of this concept of time is needed, as time is
often considered to be an illusion\cite{McTaggart08}. I find that
``illusion'' is an overly strong word here --- it implies that we are
tricked into observing something that is not, in fact, there. I would
argue that the second law of thermodynamics is also a similar
phenomenon --- it comes about from our propensity to classify systems,
for instance, according to thermodynamic variables. Yet there would be
howls of protest from my physicist colleagues if one were to assert
that the second law of thermodynamics is an illusion. What we have in
fact is two equally valid descriptions of the world --- an
``external'' one, in which time appears as a coordinate, measured with
clocks, and an ``internal'' one with present, future and past. Thus we
should really speak of psychological time as emergent, just as real,
but irreducible to the objective world of physics.

This is undoubtedly an {\em idealist} stance. Idealism contrasts with
{\em realism} by giving primacy to the subjective world of
appearances. In common with many idealist philosophies, there is a
problem to solve with the appearance of consistency between different
observer perceptions of reality. Unlike Bishop Berkeley, we need not
appeal to beneficent deity, nor unlike Kant do we need to appeal to an
unknowable {\em Noumenon} in order to explain this consistency. Rather
we appeal to what is known as the {\em Anthropic Principle}, which
effectively states that reality's appearance {\em must be} consistent
with our presence in that reality as an observer\cite{Barrow-Tipler86}.

The Anthropic Principle is in fact a tautology, if one accepts a
concrete reality (realism in other words). If the appearance of
reality were not consistent with our existence, we would conclude that
we have been imprisoned within a virtual reality of some kind, and
that bodies occupy a greater reality than the one we see. In fact, the
Anthropic Principle is not some quaint philosophical principle, but
has real scientifically testable consequences. In all such cases, the
Anthropic Principle has been confirmed, particularly in
cosmology\cite{Barrow-Tipler86}. 

In an idealistic setting, however, the anthropic principle is somewhat
of a mystery. Why shouldn't we be spending our entire lives dreaming a
reality that has nothing whatsoever to do with us as observers? If it
weren't for the consistency requirement of the anthropic principle,
the Occam's razor result would imply that we'd experience a universe
that is too simple to support life. The complexity of our observed
universe is in fact the minimum possible complexity that allows
conscious life. We can only suppose that the still to be found theory
of consciousness will require self-awareness as a necessary
requirement for consciousness. This in turn implies that we, as
observers are embedded in the reality we observe. This fundamental
closure of observer and observed, closes the ontology of bitstrings,
and cuts off the prospect of solipsism.

\section{Quantum Mechanics}

By reasoning about the appearance of reality, and of the nature of
observation, we have lead to a number of properties of consciousness we
expect a valid theory of consciousness to have. These include the
property of classifying reality (attaching meaning), psychological
time, meaning robustness and the anthropic principle. Combining
classification with time, we have {\em projection}, namely the process
of converting {\em possible} into {\em actual}. Periodically, we
observe a few more bits of our reality bitstring, and our
understanding is updated. By introducing the framework of probability
theory, one can mathematically derive a predictive theory of what is
actually observed. That predictive theory is the Hilbert space
structure of quantum mechanics, vectors in the Hilbert space evolve
according to Schr\"odingers equation and the probabilities are given by Born's
rule\cite{Standish00a}. Quantum Mechanics is simply a theory of observation!

Roy Frieden uses a slightly different formulation of statistical
inference\cite{Frieden98}. An observer trying to measure some physical
quantity, will experience an error $e$, which at minimum is bounded by
the inverse of something called {\em Fisher Information} $I$:
\begin{equation}\label{Cramer-Rao}
e^2I\geq1
\end{equation}
The ideal observer will of course attempt to observe reality in such a
way as to maximise $I$ --- such an extremum principle he calls the
{\em Principle of Extreme Physical Information}, and using it, along
with assorted symmetry principles, he is able to derive all sorts of
basic physics equations: the Klein-Gordon equation (which is the
relativistic version of Schr\"odinger's equation), Maxwell's equations
of electrodynamics and Einstein's field equations of General Relativity.

As a third possible route to quantum mechanics from observerhood,
Bruno Marchal considers the consequences of {\em computationalism},
that we as observers are equivalent to programs running on a Turing
machine (an abstract model of a computer)\cite{Marchal98}. By adopting
a theory of knowledge proposed by Plato in {\em Thaetetus}, and
translating it into logical terms, he demonstrates that knowledge
about observed reality must obey the axioms of quantum logic, a logic
which also describes the behaviour of subspaces of a Hilbert space.
Whilst the conclusions might be weaker than those derived by myself or
Frieden, it has the advantage of being an independent derivation, and
also of making use of fewer assumptions about the nature of conscious
observation that Frieden and I make.

Thus we see a reversal in the usual ontology between the observer
(``psychology'') and physics. Physics is actually emergent from
psychology, and by the Anthropic Principle, must be emergent from
physics.

\section{Conclusion}

The success in reversing the usual ontology between the observer
(``psychology'') and physics in the case of quantum mechanics leads
one to conjecture that the other pillar of 20th century physics,
Einstein's theory of relativity may also be related to properties of
the observer.

If we take this approach seriously, by showing a possible reduction of
physics to psychology, physics constrains the possible forms a theory
of consciousness might take. We have seen that the Anthropic Principle
is pivotal, and this implies self-awareness as necessary property of
consciousness. We have also seen that psychological time, is also
crucial. Even though brains are messy, asynchronous, parallel
processors, somehow a sequential process emerges to give the
appearance of psychological time. Finally, the mind attaches meaning
to data in a robust fashion. This is a selection process, analogous to
natural selection in Darwinian evolution. 

If we take a look at Daniel Dennett's {\em Consciousness Explained}\cite{Dennett91}
\footnote{For illustrative purposes, I'm not endorsing this particular theory
as a complete, final theory of consciousness.}, we see all of these
features appearing. Firstly much is made of self-awareness, which
Dennett calls {\em autostimulation}, and is what he calls a {\em good
  evolutionary trick}. Secondly, he argues that a sequential {\em von
  Neumannesque} architecture must emerge from an underlying {\em
  Pandemonium}, something he calls the {\em Joycean machine}, in
honour of James Joyce's {\em stream of consciousness}. Finally, he
argues that instead of a Cartesian theatre in which consciousness
reside, a Pandemonium of independent parallel processes compete in an
evolutionary process, that results in a series of rewritten ``drafts''
of understanding.

\bibliographystyle{plain}
\bibliography{rus}

\end{document}